\documentstyle[epsbox]{l-aa}

\begin{document}

   \thesaurus{06         
              (02.18.8;  
               08.14.1;  
               08.18.1;  
               08.13.1;  
               03.13.1)} 
   \title{Flattening modulus of a neutron star by
          rotation and magnetic field}


   \author{K. Konno \and T. Obata \and Y. Kojima
          }

   \offprints{K. Konno}

   \institute{Department of Physics, 
              Hiroshima University,
              Higashi-Hiroshima 739-8526, Japan
             }

   \date{Received }

   \maketitle

   \begin{abstract}

   We calculated the ellipticity of the deformed star due to the
   rotation or magnetic field. These two effects are compared to
   each other within general relativity. It turned out that the
   magnetic distortion is important for recently observed
   candidates of magnetars, while the magnetic effect
   can be neglected for well-known typical pulsars.

      \keywords{relativity --
                stars: neutron --
                stars: rotation --
                stars: magnetic fields --
                methods: analytical
               }
   \end{abstract}

\section{Introduction}

Observations of pulsars have been accumulated 
since the first discovery by Hewish et al.~(\cite{hew}),
and several new types of pulsars appeared with great surprise.
These observations have partially revealed
the structure and evolution of rotating neutron stars.
Their rotation periods range from $1.5\:$ms to several seconds. 
The surface magnetic fields range from $10^8 $
to $ 10^{15} \mbox{G}$.
The upper limit was recently raised by a factor $10^3 $
by the discovery of the soft gamma-ray repeaters (SGRs) 
and anomalous X-ray pulsars (AXPs)
(see e.g. Kouveliotou et al.~\cite{kouv1,kouv2};
Mereghetti \& Stella~\cite{ms}).
The new class of pulsars with the very strong magnetic field 
($B \sim 10^{14}$--$10^{15} \mbox{G}$)
are most likely candidates of {\em magnetars}
(e.g. Thompson \& Duncan~\cite{td96})
and may be worth studying further.
In the future, we may find a more extreme case, that is,
a rapidly rotating relativistic star with a strong magnetic field.

Most stars have spherically symmetric structure.
They are however deformed due to the rapid rotation 
and the strong magnetic field.
It is well known that both effects produce a flattening
equilibrium star.
We will examine them
within general relativity.
These effects are assumed to be small and 
treated as perturbations to spherically symmetric stars.
The rotational axis of pulsars does not in general coincide with
the axis of the dipole magnetic field. 
The relativistic treatments for the case is a complicated task,
because the situation is not stationary.
In this paper, however, we assume that  
the rotational effect decouples from the magnetic effect. 
Thus we consider the deformation arising from each perturbation
separately and estimate the ellipticity. 
The estimate is important to judge 
which effect dominates in the rotating magnetized stars, 
whose rotation rate and the magnetic field are in a wide range.
Our treatment is beyond the classical estimate
in the Newtonian gravity, and give
better comparison of the rotational effect
and the magnetic effect on star deformation for various pulsars.
When either of them is huge, our estimate breaks down, 
and sophisticated numerical codes are required
(see e.g. Bocquet et al.~\cite{bbgn};
Bonazzola \& Gourgoulhon \cite{bg}).

The paper is organized as follows.
In Sect.~2, we briefly review deformation of stars due to 
the rotation (Chandrasekhar \cite{chandra1}; 
Chandrasekhar \& Roberts \cite{cr})
and the magnetic field (Chandrasekhar \& Fermi \cite{cf}; 
Ferraro \cite{ferraro}; Gal'tsov et al. \cite{gtt}; 
Gal'tsov \& Tsvetkov \cite{gt}) within
Newtonian gravity. The quadrupole deformation can be 
evaluated by the ellipticity of the equilibrium shape.
In Sect.~3, we also calculate the ellipticity 
based on the general relativistic perturbation theory
(see also Hartle (\cite{hartle}) and Chandrasekhar \& Miller (\cite{cm})
for the rotational cases and Konno et al.~(\cite{kok})
for the magnetic cases).
The ellipticity can be summarized in the same form as
the Newtonian cases, but with different numerical factors.
In Sect.~4, using the ellipticity,
we compare the rotational effect
with the magnetic effect on star deformation numerically.
In this comparison, we keep the parameter range of known 
pulsars in mind.
Finally, we give concluding remarks in Sect.~5. 
Throughout the paper, we use the units in which $c=G=1$.

\section{Simple estimate of deformation}

Quadrupole deformation of an equilibrium body is characterized 
by the ellipticity $\varepsilon$,
which is defined by
\begin{equation}
\label{Def-e}
 \varepsilon = \frac{ {\rm equatorial ~radius - polar ~radius }}
    { {\rm mean ~radius } } . 
\end{equation}
For the gravitational equilibrium with uniform rotation,
the value is essentially related to the ratio of the rotational energy to 
the gravitational energy. 
In the slow rotation of a homogeneous star, 
we have a well-known result
(see e.g. Chandrasekhar \cite{chandra2}):
\begin{equation}
 \varepsilon_{\Omega} = \frac{5}{4} \frac{R^3 \Omega^2}{M},
\end{equation}
where $R$, $M$ and $\Omega$ denote the radius, mass and angular 
velocity, respectively.
For other stellar model,
the numerical factor $5/4$ should be replaced by 
an appropriate one.
For example, 
the factor is $0.76$ for the model with the polytropic 
equation of state (EOS) $p=\kappa \rho^{(n+1)/n}$ 
with index $n=1$
(see Table 1 in Chandrasekhar \& Roberts \cite{cr}).
We therefore generalize the expression with 
a dimensionless factor  $f$  as
\begin{equation}
\label{e-Newton}
 \varepsilon_{\Omega} = f  \frac{R^3 \Omega^2}{M},
\end{equation}
and will discuss how the factor $f$
depends on the stellar models.

In a similar way, the effect of the magnetic stress
is also expressed by the energy ratio of the magnetic field
to the gravitational field. Introducing a dimensionless
factor $g$,  we have 
the ellipticity $\varepsilon_{B} $ arising from 
magnetic field as
\begin{equation}
\label{e-mag}
 \varepsilon_{B} =  g \frac{\mu^2}{M^2 R^2},
\end{equation}
where $ \mu $ is the magnetic dipole moment. 
The dimensionless factor $g$, in general, depends on both the magnetic 
field configurations and the EOS. 
For example, in the case of an incompressible
fluid body with a dipole magnetic field 
treated by Ferraro (\cite{ferraro}),
we derive $g=25/2$.

\section{Relativistic calculation of deformation}

In this section, we will 
review quadrupole deformation due to the slow rotation and
weak dipole magnetic field.
The deformation can be expressed by the second-order quantities
with respect to the rotation rate or
the magnetic field strength.
In order to calculate the shape, 
we also have to calculate the space-time metric, 
which is axisymmetric stationary or static one.
The line element can be written in the form 
\begin{eqnarray}
 ds^2 & = & - \: e^{\nu (r)} 
          \left[ 1 + 2 \left\{ h_{0}(r) + h_{2}(r) 
          P_{2}\left( \cos \theta \right) 
          \right\} \right] dt^2  \nonumber \\
   & & + \: e^{\lambda (r)} \left[ 1 + \frac{2 e^{\lambda (r)}}{r} 
          \left\{ m_{0}(r) + m_{2}(r) 
          P_{2}\left( \cos \theta \right)\right\} 
          \right] dr^2 
          \nonumber \\
   & & + \: r^2 \left( 1 + 2 k_{2}(r) 
          P_{2}\left( \cos \theta \right) \right) \nonumber \\
   & & \qquad \times \;  
          \left[ d \theta^2  
          + r^2 \sin^2 \theta (d \phi - \omega (r) dt)^2 \right],
\end{eqnarray}
where $P_{2}$ is the Legendre's polynomial of degree $2$, and
$(h_{0}, h_{2}, m_{0}, m_{2}, k_{2})$ are the second-order quantities
for the rotation rate or
the magnetic field strength.
The quantity $\omega $ is the angular velocity 
acquired by an observer falling 
freely from infinity to a point $r$, and
is equal to zero for the static magnetic field deformation.

The stress-energy tensor of the perfect fluid body is  
described by 
\begin{equation}
 T^{\quad \mu}_{^{\rm (m)}\ \nu} 
 = \left( \rho + p \right) u^{\mu} u_{\nu} + p \delta^{\mu}_{\ \nu} .
\end{equation}
When we consider the magnetic field deformation, we further
take into account the stress-energy tensor arising from
the magnetic field, i.e.,
\begin{equation}
 T^{\quad \; \mu}_{^{\rm (em)}\ \nu} 
 = \frac{1}{4\pi} \left( F^{\mu \lambda}
     F_{\nu \lambda} - \frac{1}{4} F_{\sigma \lambda}
     F^{\sigma \lambda} \delta^{\mu}_{\ \nu} \right) .
\end{equation}
Solving the Einstein-Maxwell equations, we can obtain the 
second-order metric functions mentioned above.

The ellipticity of the relativistic star
can be calculated from the definition (\ref{Def-e})
as 
\begin{equation}
  \varepsilon = -\frac{3}{2} \left ( \frac{\xi_2}{r} + k_{2} 
\right ) ,
\end{equation}
where $ \xi_{2}$ represents the displacement of
quadrupole deformation.

Since the displacement of the surface can be determined 
by the hydrostatic equilibrium condition, 
the ellipticity of the slowly rotating star is
expressed as (Chandrasekhar \& Miller \cite{cm})
\begin{equation}
\label{Relp}
\varepsilon_{\Omega}
  = \frac{3}{r \nu'} h_{2} +
   \frac{r}{\nu'} e^{-\nu} (\Omega - \omega)^2 
           -\frac{3}{2}k_{2}.
\end{equation}
\begin{figure}[htbp]
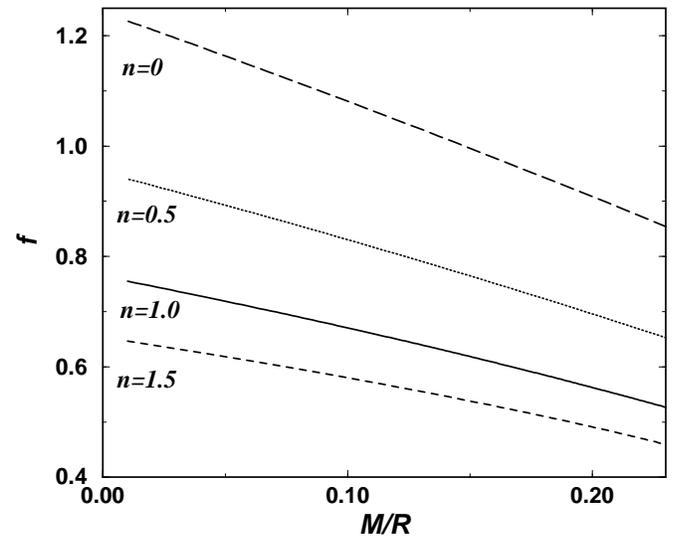

   \epsfile{file=9366.f1,width=8.8cm}
   \caption{The dimensionless factor $f$, that is, 
the ellipticity of the rotational cases 
normalized by $R^3 \Omega^2 /M$, which is plotted 
against $M/R$.}
   \label{fig1}
\end{figure}
In order to compare it with the Newtonian results,
we normalize the ellipticity in the same form as Eq.~(\ref{e-Newton}).
There are several possibilities of the normalization factors
for $M$ and $R$ in the relativistic calculation. We use
natural choices, i.e., gravitational mass for $M$ and 
circumferential radius for $R$ in this paper.
This normalization is useful to extrapolate from the 
Newtonian results. Chandrasekhar \& Miller (\cite{cm}) 
used a different normalization, which causes a
prominent peak (see Fig.~5 in their paper).
We note that the peak is due merely to a 
less convenient choice of the normalization.
The resultant dimensionless factor $f$ is
calculated for the polytropic EOS 
with $n=0, \; 0.5, \; 1, \; 1.5$.
Fig.~\ref{fig1} displays the variation of 
the dimensionless quantity $f$
with respect to the relativistic factor $M/R$.
This figure shows that the correct relativistic calculations
give smaller values of the ellipticity than those of 
the Newtonian calculations with fixed $R^3 \Omega^2 / M$.
The factor in the typical relativistic case decreases
down to 0.7 of the Newtonian case.

\begin{figure}[htbp]
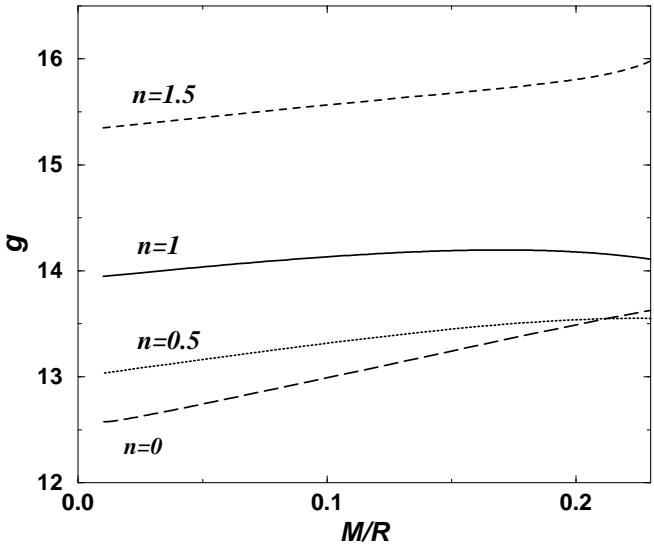

  \epsfile{file=9366.f2,width=8.8cm}
  \caption{The dimensionless factor $g$, that is, 
the ellipticity of the magnetic case 
normalized by $\mu^2 / (M^2 R^2)$, 
which is plotted against $M/R$.
(Note that the normalization of this figure is different 
from that of Konno et al.~(1999).)}
  \label{fig2}
\end{figure}
As for the weakly magnetized star,
the Lorentz force plays a role on the equilibrium.
The rotational term $\Omega - \omega$ is replaced
by the magnetic term in the ellipticity.
From the hydrostatic condition, we have
(Konno et al.~\cite{kok})
\begin{equation}
\label{Melt}
  \varepsilon_{B} = \frac{3}{r \nu'} h_{2}
        + \frac{2r}{\nu' (\rho_{0}+p_{0})} J^{\phi} A^{\phi} 
        - \frac{3}{2} k_{2} ,
\end{equation}
where the subscript `0' denotes the background quantities,
and $J^{\phi}$ and $A^{\phi}$ are the $\phi$-components
of the 4-current and 4-potential respectively.
The ellipticity can also be written in the form of Eq.~(\ref{e-mag}).
We also use the gravitational mass $M$ and circumferential 
radius $R$.
Fig.~\ref{fig2} displays the variation of the ellipticity
with respect to the relativistic factor $M/R$.
In these calculations, we have used the current distribution
\begin{equation}
 J^{\phi} \propto \rho_{0} + p_{0} ,
\end{equation}
which is a simplest case derived by the integrability condition
(see e.g. Ferraro \cite{ferraro})
and gives the direct general-relativistic extension 
of Ferraro (\cite{ferraro}).
Using the normalization, the residual factor $g$ is
almost independent of the relativistic factor $M/R$.

\section{The comparison}

In order to compare the rotational effect with the magnetic effect
on star deformation, we now consider the ratio of
$\varepsilon_{\Omega}$ to $\varepsilon_{B}$,
\begin{equation}
\label{ratio}
 \frac{\varepsilon_{\Omega}}{\varepsilon_{B}} 
  = \frac{f}{g} \: \left( \frac{R^3 \Omega^2}{M} \right) 
    \left( \frac{\mu^2}{M^2 R^2} \right)^{-1} .
\end{equation}
\begin{figure}[htbp]
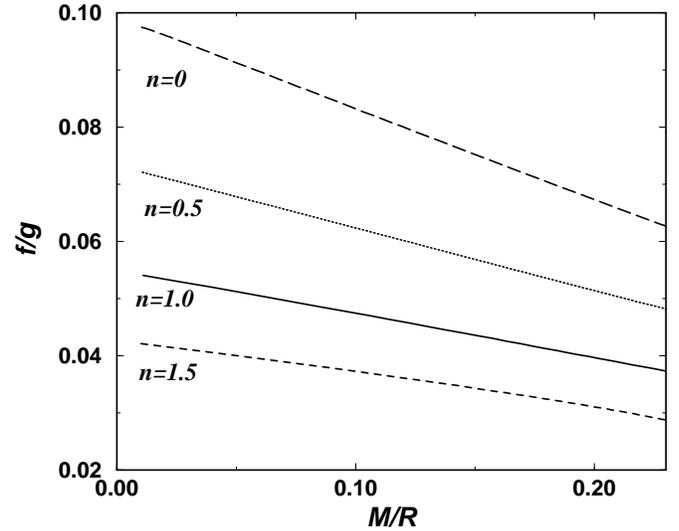

  \epsfile{file=9366.f3,width=8.8cm}
  \caption{The ratio of the two dimensionless factors,
$f/g$, plotted against $M/R$.}
  \label{fig3}
\end{figure}
First, we investigate the dependence of the ratio of the 
two dimensionless factors, $f/g$, on the relativistic factor
$M/R$. Fig.~\ref{fig3} displays this ratio.
From this figure, we can see that the true relativistic calculations
with $M/R \sim 0.2$
give smaller values of the ratio than those obtained
by the Newtonian calculations.
This fact means that the approach using the Newtonian gravity
overestimates the rotational effect.
The curves $f/g$ are approximately reproduced within 10\% 
if we use
\begin{equation}
\label{fitting}
 \frac{f}{g} \approx \frac{1-1.4M/R}{10+8n}.
\end{equation} 

\begin{figure}[htbp]
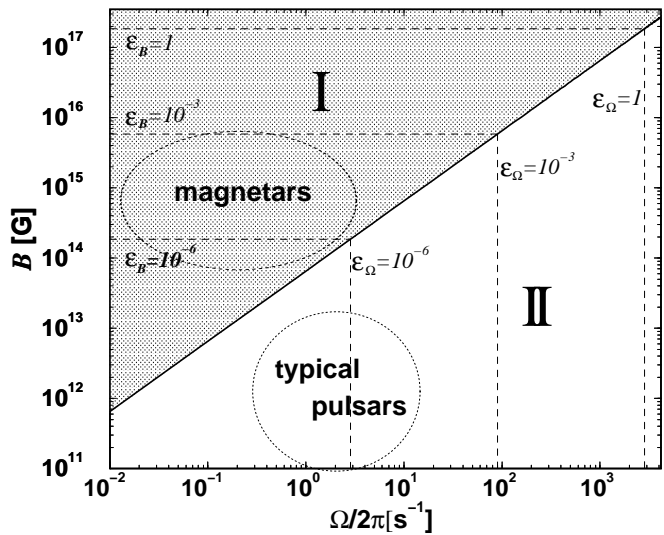

  \epsfile{file=9366.f4,width=8.8cm}
  \caption{The critical line on which $\varepsilon_{\Omega} 
= \varepsilon_{B}$ in $B$-$\Omega$ space
for $M/R = 0.2$.
This line divides the $B$-$\Omega$ space into
two regions. In the region I, the magnetic effect dominates
the rotational effect,
while in the region II vice versa.}
  \label{fig4}
\end{figure}

Next, we consider the comparison including other
parameters $\Omega$ and $\mu$ of stars.
We use the stellar model with $M=1.4 M_{\odot}$ 
and $R=10 {\rm km}$.
Fig.~\ref{fig4} displays a critical line on which
$\varepsilon_{\Omega} = \varepsilon_{B}$ and
the two regions divided by this line
in $B$-$\Omega$ space, where $B$ denotes the typical 
magnetic field strength on the surface, which is defined 
by $B=\mu / R^3$.
We have plotted only one representative line of $n=1$.
We can also derive very close results for other indices.
The critical line, in general, can be written from
Eq.~(\ref{ratio}) as
\begin{equation}
\label{c-line}
 B[\mbox{G}]
 \approx 5.3 \times 10^{13} 
 \sqrt{\frac{f}{g}} \left( \frac{M / 1.4 M_{\odot}}
 {R / 10 {\rm km}} \right)^{1/2} \Omega \: [\mbox{sec}^{-1}] ,
\end{equation} 
where it is useful to use the fitting formula Eq.~(\ref{fitting})
for $f/g$.
In the region \uppercase\expandafter{\romannumeral 1},
the magnetic effect dominates the rotational effect, i.e., 
$\varepsilon_{B} > \varepsilon_{\Omega}$,
whereas in the region \uppercase\expandafter{\romannumeral 2}
vice versa, i.e., 
$\varepsilon_{\Omega} > \varepsilon_{B}$.
From this figure, we find that objects having magnetic 
field strength $B \sim 10^{14}$--$10^{15}$G and period
$T \sim 1 \:$sec such as SGRs and AXPs belong to the 
region \uppercase\expandafter{\romannumeral 1}.
Thus, the magnetic effect overwhelms the rotational effect
for such observed candidates of magnetars.
On the other hand, well-known typical pulsars with
magnetic field strength $B \sim 10^{11}$--$10^{13} \mbox{G}$
and the period $T \sim 10^{-1}$--$1 \: \mbox{sec}$
(see e.g. Taylor et al. \cite{catalog})
obviously belong to the region 
\uppercase\expandafter{\romannumeral 2}.
Millisecond pulsars also belong to the region 
\uppercase\expandafter{\romannumeral 2}.
The magnetic deformation is neglected.

\section{Concluding remarks}

The new classes of objects, which are candidates of magnetars,
have inspired us to investigate the relation between the 
rotational effect and the magnetic effect on deformation of stars.
We have briefly reviewed the quadrupole deformation due to the
rotation and that due to the magnetic field based on previous 
studies, and compared the rotational effect with the magnetic effect
for various pulsars reported observationally.
From our investigation, we have found that the new classes of objects
such as SGRs and AXPs belong to the region in which the magnetic effect
dominates the rotational effect, while well-known typical pulsars
with magnetic field strength $10^{11}$--$10^{13}$G 
and millisecond pulsars vice versa.
Thus, the critical line on which the ellipticity arising from the
rotation equals to that arising from the magnetic field
divides the new classes and the well-known pulsars.
Once we obtain the parameters $\Omega$, $B$, $R$ and $M$ of a pulsar,
we can see whether the magnetic effect is dominant or 
the rotational effect is dominant
using Eqs.~(\ref{fitting}) and (\ref{c-line}) and Fig.~\ref{fig4}.
The deformation due to the magnetic field
may come into play in the spin-down of the magnetars. 
The spin-down history of the AXPs is rather irregular.
Recently, Melatos (\cite{melatos}) ascribed the irregularity to
the radiative precession produced by the deformation. 
At present, the fit to the observational data is not so good,
but will be improved by detailed models.
Theoretical models for the deformed stars will be required
there.

As discussed by Bonazzola \& Gourgoulhon (\cite{bg}),
the non-axisymmetric distortion is also important for 
the gravitational emission. The deformation can be 
calculated from the magnetic field, which is estimated
from the observed pulsar period and the period derivative.
The inferred amplitudes of the gravitational waves
are too small for the present known pulsars.
In the future, we might find more extreme case such as 
the early stage of rapidly rotating magnetars.

\begin{acknowledgements}
 We would like to thank Dr.~A.Y.~Potekhin for fruitful
comments and suggestions. 
\end{acknowledgements}

\end{document}